\documentclass[aps,pra,twocolumn,groupedaddress]{revtex4-2}
\usepackage{fix-cm}

\usepackage[utf8]{inputenc}

\usepackage{amsmath}
\usepackage{amssymb}
\usepackage{tikz}
\usepackage{graphicx}    
\usepackage{dcolumn}     
\usepackage{bm}          
\usepackage{xcolor}      
\usepackage{braket}      
\usepackage[normalem]{ulem}
\usetikzlibrary{matrix, positioning}
\usepackage{soul}        
\usepackage{booktabs}    
\usepackage{stfloats}    
\usepackage{placeins}    
\usepackage{lipsum}      
\usepackage{caption}
\usepackage{newtxtext, newtxmath}

\begin{document}

\preprint{APS/123-QED}

\title{Parity-Induced Exceptional Points with Localized Loss}

\author{J. R. Silva}
\email{jefferson.rocha@fis.ufal.br}

\affiliation{Instituto de F\'isica, Universidade Federal de Alagoas, Macei\'o, Brasil}

\date{\today}

\begin{abstract}
We explore the existence and stability of exceptional points (EPs) in finite waveguide arrays subject to single-site dissipation using an exact Laplace-transform framework. We show that the EP landscape is governed by a parameter-free Wronskian condition and a geometry-dependent parity effect, yielding strictly distinct spectral behaviors for even versus odd numbers of waveguides. Specifically, odd chains with edge loss suppress coalescence via a protected central mode, whereas even chains with central loss feature a universal threshold that gives rise to imaginary zero modes. These findings clarify the fundamental rules of non-Hermitian symmetry breaking in finite systems, offering a deterministic framework to predict, engineer, or avoid spectral singularities in integrated photonic lattices.

\end{abstract}

\maketitle


\section{Introduction}

The study of non-Hermitian physical systems has evolved from a fundamental mathematical curiosity into a cornerstone of modern optical engineering and classical wave dynamics~\cite{miri2019exceptional, heiss2012physics, bergholtz2021exceptional}. Unlike conservative Hermitian systems, open non-Hermitian systems explicitly account for energy exchange with the external environment through tailored gain and dissipation. This paradigm shift has enabled a plethora of counterintuitive phenomena, including unidirectional invisibility~\cite{lin2011unidirectional, feng2013experimental}, single-mode lasers with enhanced stability~\cite{feng2014single, hodaei2014parity}, and anomalous scattering behavior. Central to these advances are Exceptional Points (EPs)~\cite{klauck2025crossing, bergholtz2021exceptional}, spectral singularities where two or more complex eigenvalues and their corresponding eigenvectors simultaneously coalesce. This defective degeneracy renders the effective Hamiltonian non-diagonalizable and induces an extreme topological sensitivity to external perturbations, establishing EPs as a powerful tool for high-precision optical sensing~\cite{wiersig2020review}.

Coupled waveguide arrays, described by tight-binding lattice models, serve as an ideal platform for exploring non-Hermitian dynamics due to the precise experimental control available over inter-guide coupling and localized loss mechanisms. However, existing literature has predominantly focused on high-symmetry regimes, such as parity-time ($\mathcal{PT}$)-symmetric dimers ($N=2$), trimers, or translationally invariant infinite lattices~\cite{ruter2010observation, ge2015parity, xue2017}. While these idealizations provide valuable baseline physics, practical integrated photonic circuits are inherently finite\cite{xiao2020non} and frequently subject to localized dissipative defects, whether introduced intentionally for mode filtering~\cite{feng2017non} or arising from local scattering losses\cite{stegmaier2021topological}. The symmetry breaking caused by a single dissipative impurity in a finite chain introduces complex boundary interactions and finite-size interference effects that cannot be captured by infinite-lattice approximations.

Crucially, localized dissipation does not merely act as an energy sink; it reshapes the global spectral topology of the lattice. Depending on the loss magnitude, a localized loss site can trigger two distinct physical regimes. At moderate dissipation rates, it drives eigenvalue coalescence and phase transitions. At high dissipation rates, however, it acts as an effective spatial potential barrier due to the dissipative Quantum Zeno effect~\cite{snizhko2020quantum, facchi2008quantum}, effectively partitioning the chain into decoupled sublattices\cite{ding2022non, ashida2020non} and inducing a transition between mode absorption and reflection. Furthermore, the interplay between the loss location $j$ and the total system size $N$ gives rise to strong discrete geometric constraints, such as parity effects, which remain under-explored in discrete non-Hermitian systems\cite{lin2019symmetry}.

From an analytical standpoint, determining the location of EPs in arbitrary $N$-site lattices usually relies on numerical matrix diagonalization, which obscures the underlying algebraic structure\cite{yang2020non}. In this work, we overcome this limitation. We demonstrate that the spectral landscape of a finite chain with single-site loss is dictated by a parameter-free Wronskian determinant condition. This exact formulation reveals that the existence and stability of EPs are strictly governed by a parity effect: odd-sized chains with edge loss suppress EP coalescence due to a topologically protected zero mode anchored at the lattice center, whereas even-sized chains with central loss exhibit a universal coupling threshold of $\sigma = 2\kappa$, generating purely imaginary zero modes.

Furthermore, the presence of protected modes, often referred to as dark modes, demonstrates how the spatial symmetry of a photonic lattice can shield light from medium dissipation\cite{yang2020non}. By relying on destructive interference precisely at the loss-affected waveguides, these modes become completely isolated from attenuation, ensuring the survival and stability of the average photon number during propagation\cite{weidemann2020topological}. This immunity to dissipation transforms the lattice geometry itself into a natural shield, enabling a lossless transport channel for photonic information\cite{longhi2022non}.

The article is structured as follows: Section I introduces the effective non-Hermitian Hamiltonian for $N$ coupled waveguides with localized dissipation, deriving the exact spatial transfer matrix via Fibonacci polynomials. Section II systematically maps the exceptional point landscape using a Wronskian criterion, emphasizing the role of spatial parity and loss location. Section III investigates the formation of protected modes, demonstrating how the geometric proportions of the lattice govern topological shielding and asymptotic photon survival. In Section IV, we extend the lattice with an auxiliary waveguide, illustrating its function as a tunable Lorentzian reservoir that generates third-order exceptional points. Concluding remarks and physical implications for optical design are presented in Section V.

\section{Hamiltonian and Heisenberg equations solution}

Initially, we consider a one-dimensional array of single-mode coupled waveguides, $G_1$, $G_2$, $\cdots$, $G_{N-1}$, $G_N$, where, for some $j$, $G_j$ has a loss to the environment that can be mapped by a Markovian reservoir.

\begin{center}
  \includegraphics[width=0.48\textwidth]{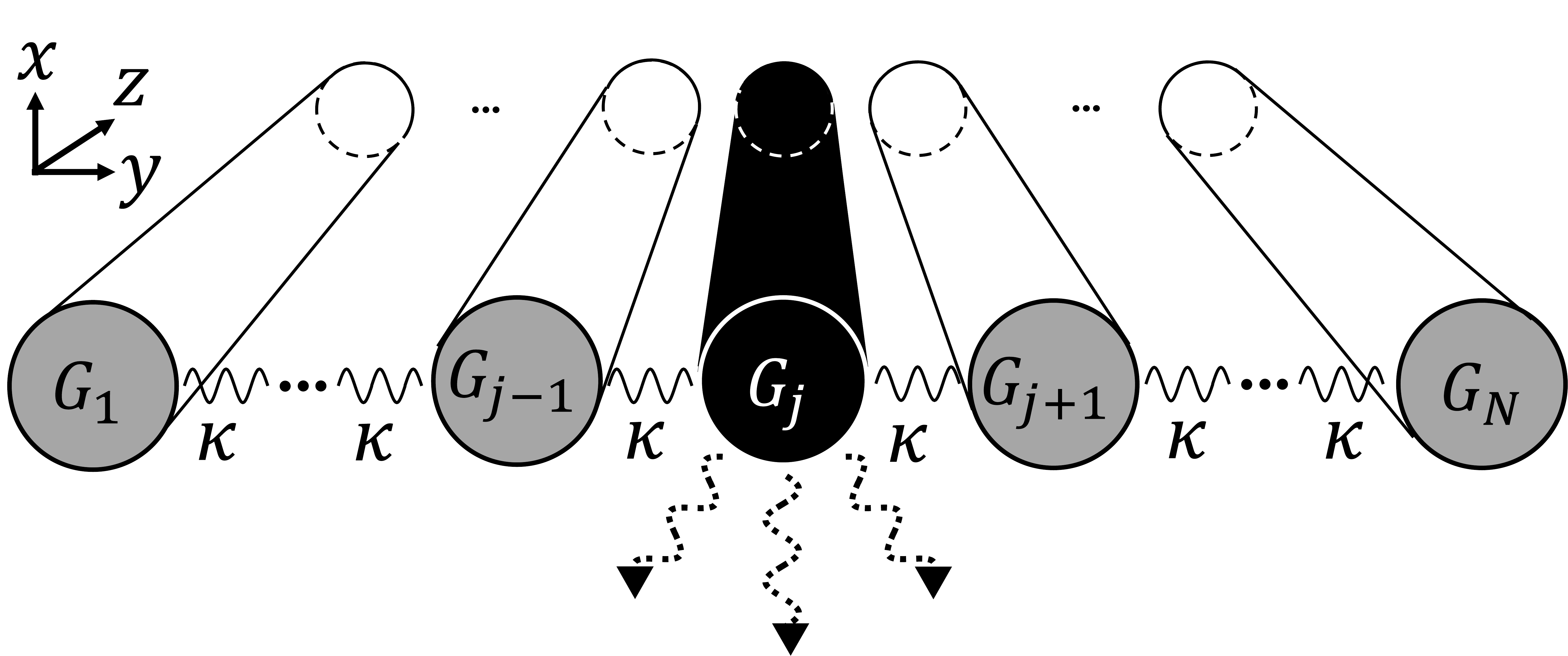}
  \captionsetup{hypcap=false}
  \captionof{figure}{Sequence with $N$ coupling waveguides, with coupling constant $\kappa$, where the waveguide $G_j$ has Markovian loss indicated by the dashed arrows.}
  \label{model}
\end{center}

The Hamiltonian of the set of coupled guides observed in figure \ref{model}, in the RWA approximation, is described by $H_G$ in the presence of the environment

\begin{equation}
    H_G = \beta \sum_{m=1}^N a_m^{\dagger}a_m\,+\kappa\sum_{0<l<N}(a_l^{\dagger}a_{l+1}+a_la_{l+1}^{\dagger}) ,
\end{equation}

where $\beta$ and $\kappa$ denote the propagation constants and the mutual coupling between the waveguides, respectively, and $a_m$ and $a_m^{\dagger}$ represent the annihilation and creation operators for waveguide $m$ in the standard Schrödinger picture. The effective Hamiltonian, derived by integrating out the reservoir degrees of freedom, i.e., assuming the Markovian limit and performing the trace over the environment, is given by

\begin{equation}
    H_{\text{eff}} = H_G-i\sigma a_j^{\dagger}a_j,
\end{equation}

with $\sigma$ being a constant associated with the loss rate. 

In the Heisenberg picture, we write the spatial evolution equations for the annihilation operators of each waveguide in the system using the coupled equation

\begin{equation}\label{syst}
\frac{d}{dz}
\begin{tikzpicture}[baseline={(current bounding box.center)}]
    \matrix (V) [matrix of math nodes, left delimiter={[}, right delimiter={]},
        column sep=0.5em, 
        row sep=0.1em 
        ] 
    {
        a_1 \\ 
        |(V-centro)| \hphantom{a_N} \\ 
        a_N \\ 
    };
    
    \draw[thick] (V-1-1.south) -- (V-3-1.north);
\end{tikzpicture}
=
-i
\begin{tikzpicture}[baseline={(current bounding box.center)}, scale=1.0]
    \matrix (M) [matrix of math nodes, left delimiter={[}, right delimiter={]},
        nodes={inner sep=1pt},
        column sep=1.8em, row sep=1.5em,
        ampersand replacement=\&]
    {
        \beta_1 \& \kappa \& \, \\
        \kappa \& \, \& \kappa \\
        \, \& \kappa \& \beta_N \\
    };
    
    \begin{scope}[thick, black]
        \draw (M-1-1.south east) -- (M-3-3.north west); 
        
        \draw (M-1-2.south east) -- (M-2-3.north west);

        \draw (M-2-1.south east) -- (M-3-2.north west);
    \end{scope}
\end{tikzpicture}
\begin{tikzpicture}[baseline={(current bounding box.center)}]
    \matrix (V) [matrix of math nodes, left delimiter={[}, right delimiter={]},
        column sep=0.5em, 
        row sep=0.1em 
        ] 
    {
        a_1 \\ 
        |(V-centro)| \hphantom{a_N} \\ 
        a_N \\ 
    };
    
    \draw[thick] (V-1-1.south) -- (V-3-1.north);
\end{tikzpicture}
\end{equation}
where $\beta_m=\beta-i\delta_{jm}\sigma$, indicating the presence of Markovian loss in waveguide $G_j$. To solve the system given by $da/dz=-iMa$, we move to the Laplace domain, variable $s$, and transform back to the spatial domain $z$ via $a(z)=\mathcal{L}^{-1}\{(s+iM)^{-1}\}a(0)$.

From the perspective of the tridiagonal matrix in \eqref{syst}, the proposed $N$-waveguide system has eigenvalues with associated eigenspaces of dimension 1. This means that eigenvalue degeneracy directly implies eigenvector coalescence.

The spatial evolution of light intensity in a given waveguide $k$ is quantified by its average photon number, calculated via the expectation value of the number operator. Mathematically, this quantity is expressed as $n_k(z) = \langle \psi(0) \vert{} a_k^\dagger(z) a_k(z) \vert{} \psi(0) \rangle$, where $\vert{}\psi(0)\rangle$ represents the initial state of the system at the injection point. Thus, each entry of $a(z)$ takes the form

\begin{equation}\label{aza}
    a_m(z)=e^{-i\beta z}\sum_{k=1}^Nf_{mk}^{(j)}(z)a_k(0).
\end{equation}

The squared modulus of the amplitude $f_{mk}^{(j)}(z)$ physically represents the relative fraction of the average photon number in waveguide $G_m$ at position $z$ when photons are launched into waveguide $G_k$ at $z=0$, with loss present in waveguide $G_j$. In the lossless case ($\sigma=0$), $f^{(j)}\equiv f$ takes the well-known form

\begin{equation}
    f_{mk}(z)=\frac{2}{N+1}\sum_{n=1}^N\sin\left(\frac{kn\pi}{N+1}\right)\sin\left(\frac{mn\pi}{N+1}\right)e^{-i\lambda_nz}
\end{equation}
where $\lambda_n=2\kappa \cos\left[n\pi/(N+1)\right]$ are the eigenvalues of the Hermitian Hamiltonian; thus, the periodic nature of these functions is observed only for $N=2$ and $N=3$, since the projection of the angular spectrum is equispaced. The presence of sinusoidal terms in this equation precisely reflects the imposition of boundary conditions, where the waveguides at the ends of the chain act as rigid walls. This physical constraint gives rise to transverse interference between incident and reflected waves, forming transverse standing-wave patterns that define the system's coupled modes. Simultaneously, the exponential term $e^{-i\lambda_nz}$ acts as the plane-wave component, ensuring continuous, unobstructed energy propagation along the $z$-axis. 

In the presence of loss in waveguide $G_j$, we present the coefficients of Eq. \eqref{aza} in the Laplace domain with $m\ge k$

\begin{equation}\label{dis}
\begin{split}
    f_{mk}^{(j)}(z)
=& i^{k-m} \, \mathcal{L}^{-1}\\
&
\begin{cases}
\displaystyle
\theta_{N-m}
\dfrac{\theta_{k-1}+(\sigma/\kappa)\,\theta_{j-1}\theta_{k-j-1}}
{\theta_N+(\sigma/\kappa)\,\theta_{j-1}\theta_{N-j}},
& m > j \\[1.2em]
\displaystyle
\theta_{k-1}
\dfrac{\theta_{N-m}+(\sigma/\kappa)\,\theta_{j-m-1}\theta_{N-j}}
{\theta_N+(\sigma/\kappa)\,\theta_{j-1}\theta_{N-j}},
& m \le j
\end{cases}
,
\end{split}
\end{equation}
where $\theta_m=\theta_m(s/\kappa)$ denotes the Fibonacci polynomials, defined by $\theta_0(x)=1$, $\theta_1(x)=x$, and the recurrence relation $\theta_m(x)=x\theta_{m-1}(x)+\theta_{m-2}(x)$ for higher orders

\begin{equation}
    \begin{split}
        \theta_2(x)=&\,\,x^2+1\\
        \theta_3(x)=&\,\,x^3+2x\\
        \theta_4(x)=&\,\,x^4+3x^2+1\\
        \vdots &\\
        \theta_N(x)=&\sum_{m=0}^{\lfloor N/2 \rfloor}\binom{N-m}{m}x^{N-2m}
    \end{split}
\end{equation}

The common denominator $\theta_N + (\sigma/\kappa)\theta_{j-1}\theta_{N-j}$ defines the global characteristic equation of the perturbed system. The product $\theta_{j-1}\theta_{N-j}$ elegantly captures the physics of the problem: it reflects the decoupling induced by the loss defect, illustrating the splitting of the original chain of size $N$ into two independent sub-chains of sizes $j-1$ and $N-j$. This mathematical structure dictates the new resonance poles, highlighting how a local loss globally restructures the entire network. We observe that Eq. \eqref{dis}, at $\sigma=0$, has the denominator $\theta_N$ whose roots are $s=i\lambda_n$, thus recovering the Hermitian case.

\section{protected modes}

The long-term survival of photons in dissipative photonic lattices is intimately linked to the formation of dark modes, eigenstates of the original Hermitian lattice that exhibit a destructive interference node exactly at the dissipative waveguide. In an $N$-dimensional lattice with localized loss at site $j$, the condition for a fundamental mode $m$ to be immune to attenuation requires its spatial amplitude at that specific site to be strictly zero; that is, $\sin(m \pi j/(N+1)) = 0$.

To systematize this physical condition universally across any geometry, we evaluate the greatest common divisor $g=\gcd(N+1, j)$. This geometric variable rigorously dictates the shielding properties of the system. The exact number of protected modes supported by the lattice is given by $g-1$. Consequently, topological shielding of the light requires that $g > 1$. If $N+1$ and $j$ are coprime ($g=1$), the lattice will not support any protected modes, and the entire energy of the system will invariably dissipate in the long-propagation limit, regardless of the loss magnitude.

When the dynamics are initiated by injecting light into the first waveguide, defined by the initial state $\vert{}\psi(0)\rangle = \vert{}1\rangle$, the fraction of light intensity that resists dissipation is determined by the projection of this edge state onto the orthogonal basis of the protected modes. Each protected mode $m_l$ (where $m_l = l(N+1)/g$ for $l = 1, \dots, g-1$) captures a fraction of the initially injected light equivalent to $\langle \phi_{m_l} \vert{} \psi(0) \rangle$. The total surviving intensity in the asymptotic limit ($kz \to \infty$) is the sum of the isolated spectral weights of these modes:

\begin{equation}
    \begin{split}
       T_{\infty} &= \sum_{1\le l<g}\left\vert{} \langle \phi_{m_l} \vert{} \psi(0) \rangle \right\vert{}^2 \\&= \frac{2}{N+1} \sum_{1\le l<g} \sin^2\left(\frac{l \pi}{g}\right)\\
    &=\frac{g}{N+1} 
    \end{split}
\end{equation}

This expression demonstrates that photonic survival is inherently quantized by the geometric proportions of the lattice. For instance, in an array of $N=7$ waveguides with a lossy site at $j=6$, we obtain $g = \gcd(8,6) = 2$. This guarantees the existence of a single protected mode and yields an exact asymptotic survival of $2/8=0.25$ of the initially edge-injected photons.

Asymptotic transmission in one-dimensional lattices with localized dissipation is fundamentally governed by spatial coherence and wave interference within the system's protected modes. Exciting both edges of the lattice simultaneously introduces opposing wavefronts; their resulting interference dictates the asymptotic energy coupling based on the intrinsic geometric parity of each surviving mode, as illustrated in Figure \ref{Njz}.

\begin{center}
  \includegraphics[width=0.48\textwidth]{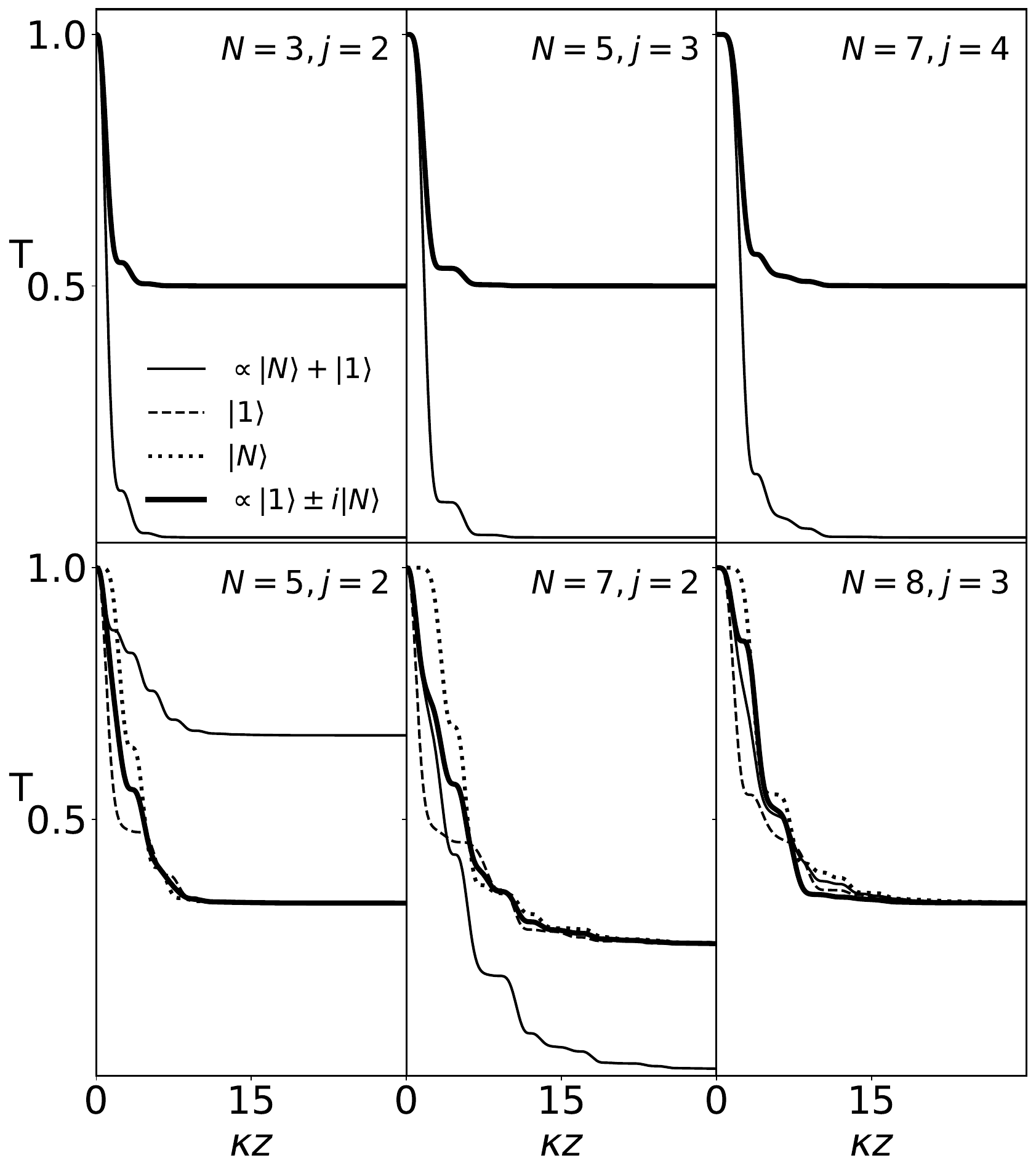}
  \captionsetup{hypcap=false}
  \captionof{figure}{Transmission as a function of propagation distance for configurations featuring protected modes, with different initial conditions.}
\label{Njz}
\end{center}

The injection of perfectly symmetric ($\propto \vert{}1\rangle + \vert{}N\rangle$) or antisymmetric ($\propto \vert{}1\rangle - \vert{}N\rangle$) states acts as a physical parity filter within the lattice. The asymptotic behavior is divided into three distinct interference regimes.

\begin{itemize}\item \textbf{Total Suppression}: Occurs when the parity of the injected state is purely orthogonal to that of all available protected modes. Destructive interference prevents the occupation of the surviving modes, forcing all energy to couple into the lossy modes and completely nullifying the transmission.
\item \textbf{Amplification}: Arises when the injected state possesses an identical symmetry to that of the protected modes. The interference is globally constructive, allowing the lattice to capture twice as much energy compared to a single-edge injection.

\item \textbf{Conservation}: Manifests in specific configurations that support a perfectly balanced number of even and odd protected modes (such as $N=8$, $j=3$). In these cases, the energy gain generated by constructive interference in one half of the subspace is exactly and globally compensated by the annihilation of the other half, stabilizing the overall transmission.\end{itemize}

The injection of superposition states with imaginary components ($\propto \vert{}1\rangle \pm i\vert{}N\rangle$) introduces a strict initial phase shift of $\pi/2$ between the edge oscillators. Physically, the intensity of the wave interference depends on the $\cos(\Delta\phi)$ term. With a phase difference of $\pi/2$, this cross-interference term completely vanishes. The absence of constructive or destructive overlap prevents the waves from reinforcing or canceling their amplitudes at the lattice nodes. Consequently, the system processes the energy as an incoherent superposition of two independent edge excitations. This shields the dynamics from the geometric constraints of the waveguide, ensuring that the asymptotic transmission retains exactly the same value as a single-edge photon injection, regardless of the internal parity configuration.

\section{Exceptional Points and Geometric Parity Effects}

Equation \eqref{dis} can be analytically written in position space using the partial fraction decomposition method, provided the roots of the polynomial $\theta_N+(\sigma/\kappa)\theta_{j-1}\theta_{N-j}$ are known. EPs arise when this polynomial exhibits repeated roots. In this instance, we evaluate the system formed by the zeros of the described polynomial and its derivative, or equivalently, where $\theta_N$ and $\theta_{j-1}\theta_{N-j}$ are pointwise linearly dependent. The Wronskian follows from this discussion:
\begin{equation}
    W_{N,j}=(\theta_{j-1}\theta_{N-j})\theta_N'-(\theta_{j-1}\theta_{N-j})'\theta_N
\end{equation}

The Wronskian possesses no free parameters other than $N$ and $j$, allowing for the rapid detection of the critical rates $\sigma$ at which exceptional points occur. The parity operation yields a physically equivalent system, an observation that follows from the index symmetry $f^{(N+1-j)}_{N+1-k,N+1-m}=f^{(j)}_{m,k}$. By virtue of this symmetry, we may restrict our analysis to cases where $j\le\lceil N/2\rceil$. Figure \ref{coales} displays the critical values of $\sigma$ for given $N$ and $j$. 

\begin{center}
  \includegraphics[width=0.48\textwidth]{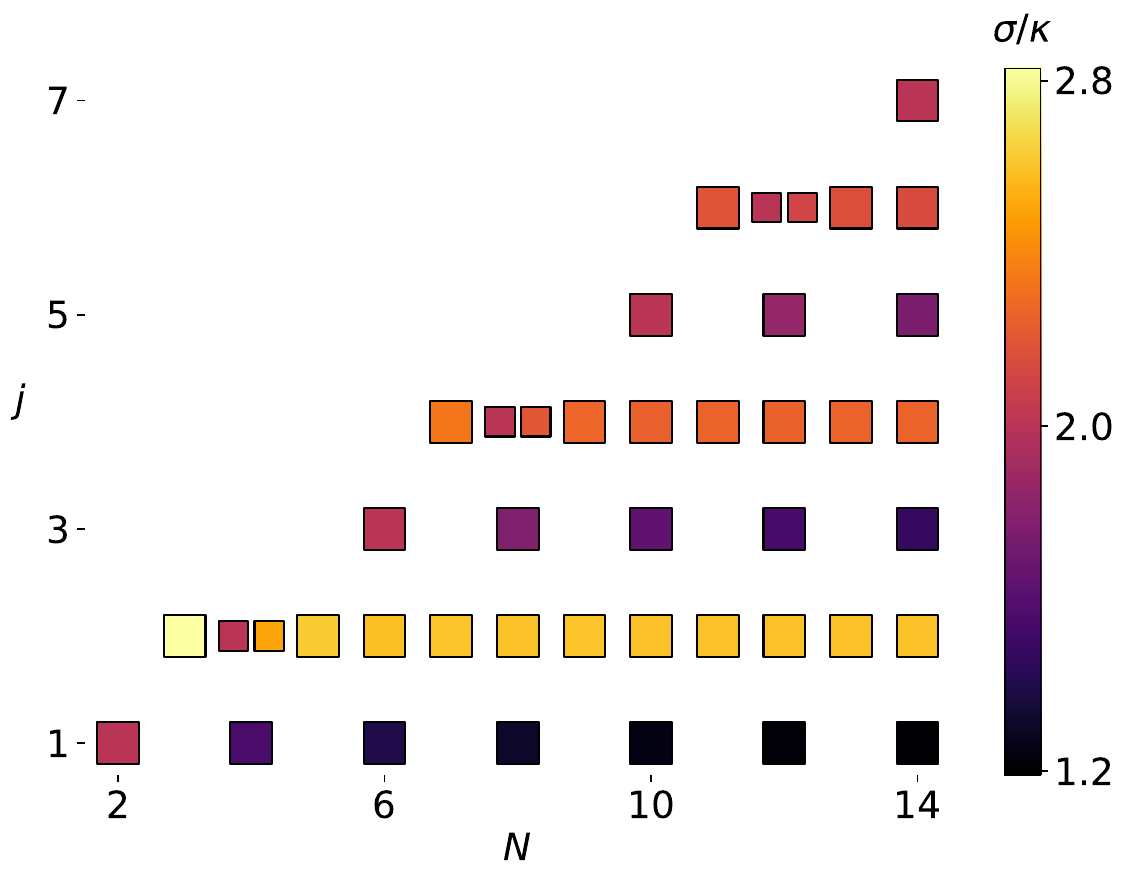}
  \captionsetup{hypcap=false}
  \captionof{figure}{Critical loss parameter $\sigma/\kappa$ for a sequence of $N$ guides with loss in guide $G_j$, with $j\le\lceil N/2\rceil$.}
\label{coales}
\end{center}

The existence of exceptional points (EPs) in this system is fundamentally governed by the parities of the total number of waveguides, $N$, and the loss position, $j$. Notably, when both $j$ and $N$ are odd, the system does not admit a critical loss parameter $\sigma$. Conversely, when $j$ is even, the system exhibits an EP. A particularly interesting phenomenon occurs for even $N$ when the loss is applied to one of the central waveguides: the critical loss parameter becomes exactly $\sigma=2\kappa$, a threshold directly inherited from the dimer case\cite{burke2020non,longhi2018quantum}.

The underlying physics becomes clearer when analyzing the boundary case, $j=1$. For odd $N$, the absence of EPs is justified by the existence of a zero mode that keeps its real part anchored at the center of the spectrum. While the lateral modes shift as loss increases, this central mode acts as a topological barrier. Because it lacks a conjugate partner, the typical coalescence required to form an EP is prevented. If $N$ is even, however, the dynamics change: the central eigenvalues merge to generate a zero mode strictly confined to the imaginary axis, which keeps the remaining modes separated without coalescing. For arbitrary internal positions ($j \ne 1$), the lattice can be conceptualized as a combination of two sublattices, $[1,j]$ and $[j,N]$, effectively coupled by the dissipative site at their mutual extremity. Under this framework, the emergence of exceptional points is guaranteed provided that the size of at least one of the resulting adjacent subchains ($j-1$ or $N-j$) is even.

Observe that a system composed of an even number of waveguides with $\sigma=2\kappa$ and $j=N/2$ renders the system matrix non-diagonalizable; this choice fuses eigenvalues and eigenvectors in pairs. Note the common denominator of the amplitudes in the Laplace domain in Eq. \eqref{dis}: $\sigma/\kappa=2$ makes them perfect squares pointwise due to the relation: $\theta_N+2\theta_{N/2}\theta_{N/2-1}=(\theta_{N/2}+\theta_{N/2-1})^2$.

For $\sigma$ slightly larger, the eigenvalues separate again, except when one of the collapses occurs generating a zero mode. When $N$ is even but not a multiple of 4, for example, the generation of the zero mode occurs immediately after $\sigma=2\kappa$. Observe in Figure \ref{N=6} the eigenvalues $\lambda\approx\pm1.30714-0.21508i$ and the formation of the zero mode at $\lambda\approx-0.56984i$.

\begin{center}
  \includegraphics[width=0.48\textwidth]{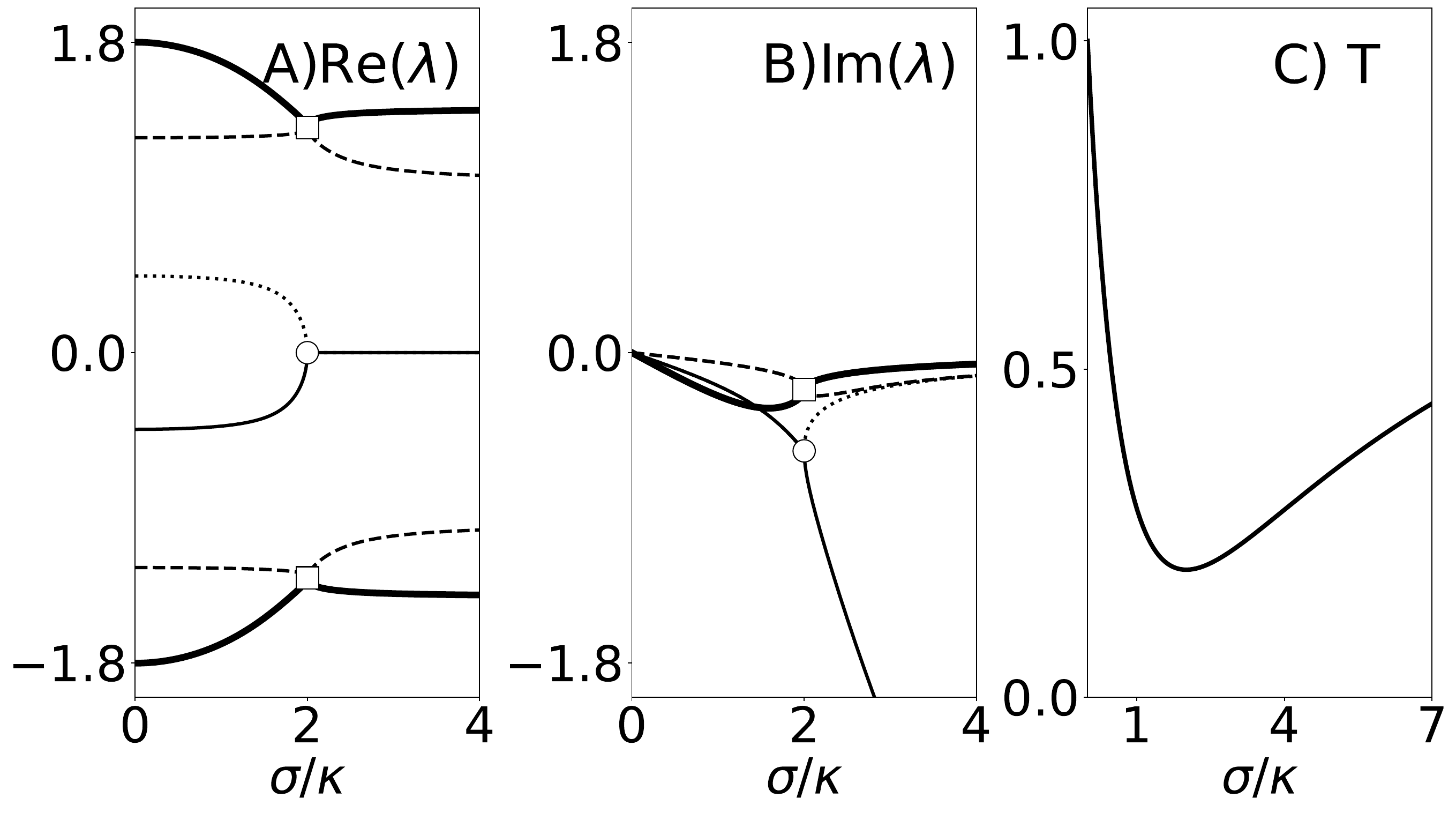} \captionsetup{hypcap=false}
  \captionof{figure}{Eigenvalue trajectories versus the loss ratio $\sigma/\kappa$ for a chain with $N=6$ and loss at site $j=3$. Panels (A) and (B) display the real and imaginary parts of the spectrum, respectively. Panel (C) shows the transmission as a function of the loss for $\kappa z=6$.}
\label{N=6}
\end{center}

When $N$ is a multiple of 4, a new EP emerges at $\sigma > 2\kappa$, yielding a zero mode, as shown in Fig. \ref{N=4}. We observe this phenomenon in panel (A) for the canonical case $N=4$, where the eigenvalues are $\lambda = i\exp(\pm 2\pi i/3)$ at $\sigma = 2\kappa$, while at $\sigma = 2.5\kappa$, they become $i$ and $[\pm\sqrt{15}+i]/4$.

\begin{center}
  \includegraphics[width=0.48\textwidth]{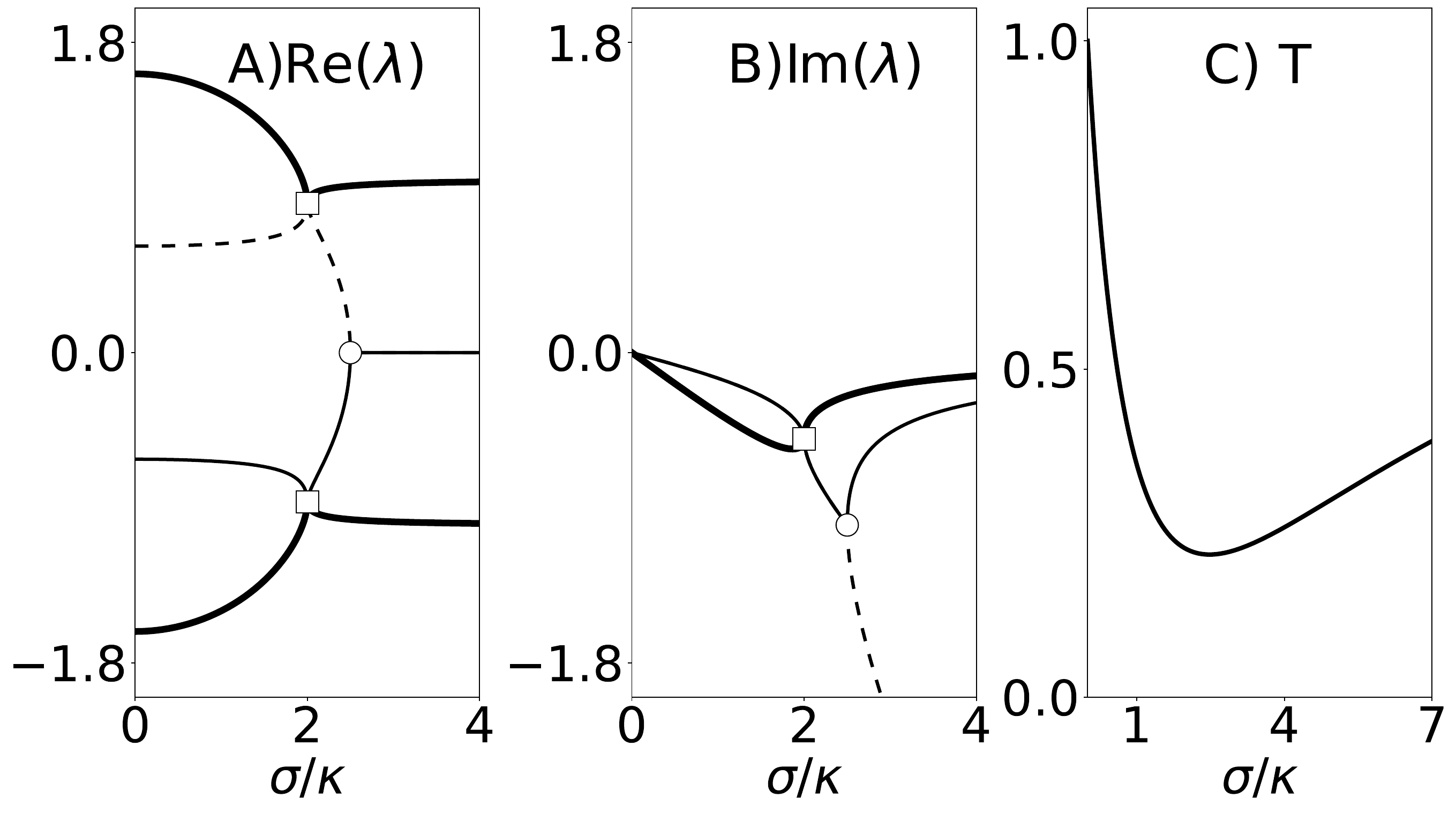}
  \captionsetup{hypcap=false}
  \captionof{figure}{Eigenvalue trajectories versus the loss ratio $\sigma/\kappa$ for a chain with $N=4$ and loss at site $j=2$. Panels (A) and (B) display the real and imaginary parts of the spectrum, respectively. Panel (C) shows the transmission as a function of the loss for $\kappa z=4$.}
\label{N=4}
\end{center}

Localized loss serves as a mechanism for topological selection. At the boundary, it effectively reduces the system dimensionality ($N \to N-1$), compelling a global spectral redistribution that can reveal a topological zero mode. Conversely, bulk loss functions as an infinite potential barrier, splitting the chain into two independent subsystems where the existence of EPs or zero modes is dictated solely by the local geometry of the isolated sub-lattices.

Both Figure \ref{N=4}C) and Figure \ref{N=6}C) demonstrate the phenomenon of loss-induced transparency, a behavior in which local dissipation paradoxically promotes transmission. Rather than simply attenuating the wave, the abrupt introduction of losses modifies the system's coupled modes, creating a barrier that isolates the defect. This blocking effect prevents the beam from interacting with the dissipative waveguide, ensuring that larger portions of light remain spatially confined and continue propagating through the lossless waveguides, effectively bypassing the region of strong absorption.

\section{SYSTEM EXTENSION: LORENTZIAN Reservoir}

To elucidate the EP dynamics, the system is extended by coupling an extra waveguide to guide $G_N$, yielding $N+1$ elements. This auxiliary guide possesses a coupling constant $\alpha \neq \kappa$ and a loss rate $\sigma = \gamma/2$. In Laplace space, the determinant reads $|s+iM|=(s+\gamma/2)\theta_N+\alpha^2\theta_{N-1}$. The exceptional points of this extended system are parameterized by the double roots of $(x-\gamma/2\kappa)\theta_N+\frac{\alpha^2}{\kappa^2}\theta_{N-1}$, under the variable change $x=-s/\kappa$. The loss rate and the coupling strength are thus described by

\begin{equation}\label{paraquem}
\left\{
\begin{aligned}
\frac{\gamma}{\kappa} &= 2x + \frac{2 \, \theta_N(x) \, \theta_{N-1}(x)}{\theta_N'(x)\, \theta_{N-1}(x) - \theta_N(x)\, \theta_{N-1}'(x)} \\
\frac{\alpha}{\kappa} &= \frac{\theta_N(x)}{\sqrt{\theta_N'(x)\, \theta_{N-1}(x) - \theta_N(x)\, \theta_{N-1}'(x)}}
\end{aligned}
\right.,
\end{equation}
Defining $W=\theta_N'\theta_{N-1}-\theta_N\theta_{N-1}'$, the parametrizations satisfy $x>0$ for odd $N$ and $x> W^{-1}(0)$ for even $N$.

The curve described by Eq. \eqref{paraquem} traces the locus of repeated eigenvalues; Fig. \ref{lorentz} displays this boundary, which separates the eigenvalue classes\cite{cavaliere2025dynamical, silva2025anomalous}.

\begin{center}
  \includegraphics[width=0.48\textwidth]{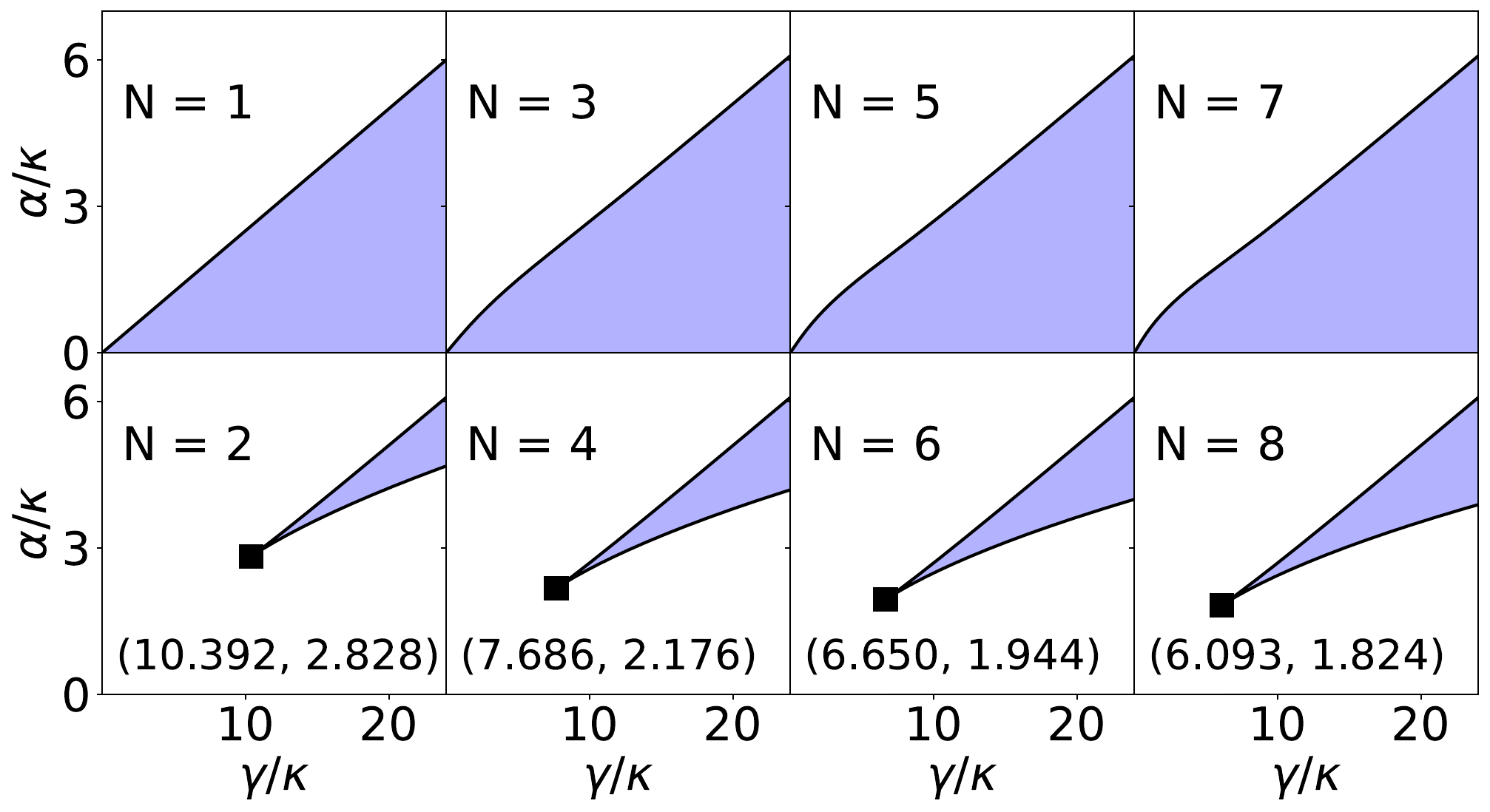}
  \captionsetup{hypcap=false}
  \captionof{figure}{Regions delineating the eigenvalue spectrum's nature. The purple shaded region denotes the environment with some purely imaginary eigenvalues.}
\label{lorentz}
\end{center}

For $N=1$, eigenvector coalescence occurs along the line $\alpha=\gamma/4$. All odd-$N$ cases exhibit a partition of solution regions that appear as deformations of this initial case. In the horizontal line $\alpha=\kappa$, the intersection with the coalescence curve occurs at increasingly smaller $\gamma$ values as $N$ increases. This decrease is observed in panel $j=1$ of Fig. \ref{coales}.

The even-$N$ cases are more peculiar, featuring a coalescence point of three eigenvalues (EP3). This triple root occurs when the parameterization in Eq. \eqref{paraquem} reaches a critical point, satisfying $\frac{d\alpha}{dx}=0=\frac{d\gamma}{dx}$. With some algebra, this corresponds to $x=\left[\frac{d^2}{dx^2}\frac{\theta_{N-1}}{\theta_N}\right]\,^{-1}(0)$. We can tabulate these values and observe an apparent decrease for this special point.

\begin{center}
  \captionsetup{hypcap=false}
  \captionof{table}{Simultaneous coalescence points of three eigenvalues.}
  \label{tab:triple_ep}
  \begin{tabular}{c c c c}
  \hline
  $N$ & $x$ & $\gamma/\kappa$ & $\alpha/\kappa$ \\ 
  \hline
  2  & $1.7320508076$ & $10.392304845413$ & $2.82842712475$ \\
  4  & $1.1631827081$ & $7.685672761047$  & $2.17568084056$ \\
  6  & $0.9027678865$ & $6.649939740468$  & $1.9439583753$ \\
  8  & $0.7477721104$ & $6.092654106580$  & $1.82414907376$ \\
  10 & $0.6431138119$ & $5.740902957312$  & $1.75050496066$ \\
  12 & $0.5669207624$ & $5.497106462282$  & $1.70044050968$ \\
  14 & $0.5085871334$ & $5.317379517210$  & $1.66407883274$ \\
  16 & $0.4622814204$ & $5.178951502979$  & $1.63640404897$ \\
  18 & $0.4245053279$ & $5.068784002605$  & $1.61459330451$\\
  \hline
  \end{tabular}
\end{center}

In the limit of arbitrarily large $N$, the coordinate $x$ corresponding to the triple eigenvalue merger tends to zero. Through an asymptotic expansion of the Fibonacci polynomials in Eq. \eqref{paraquem}, this yields the limiting values $\gamma \to 4\kappa$ and $\alpha \to \sqrt{2}\kappa$. These asymptotic results underpin the physical interpretation of the system beyond a simple linear chain.

By restricting our analysis to the $N$ functional elements, we observe that the additional dissipative waveguide effectively acts as a tunable Lorentzian reservoir\cite{silva2026non, silva2025anomalous}, where $\gamma$ represents the full width at half maximum (FWHM) of the distribution\cite{mouloudakis2022arbitrary, silva2025anomalous}. Unlike a passive drain that merely absorbs energy, this reservoir modifies the main chain's spectral topology, enabling the manipulation of Exceptional Points and the transition between absorption and reflection regimes.

This reservoir perspective elucidates the mechanism behind the dissipative Zeno effect\cite{snizhko2020quantum, facchi2008quantum}: the mismatch imposed by the Lorentzian reservoir creates an effective barrier, confining the dynamics within the original $N$ waveguides. Therefore, our study confirms that the controlled introduction of non-Hermiticity via artificial reservoirs offers a robust degree of freedom for spectral engineering, allowing finite waveguide chains to exhibit complex behaviors typically associated with much larger or continuous systems.

\section{conclusions}
In this work, we have presented an exact analytical investigation of the formation, spectral distribution, and stability of exceptional points  in finite waveguide arrays subject to localized single-site dissipation. We derived closed-form expressions for the spatial transfer amplitudes and established a parameter-free Wronskian determinant condition that precisely dictates the critical loss rates required for eigenvalue coalescence.

Our exact formalism reveals that the spectral landscape of finite non-Hermitian lattices is fundamentally governed by a spatial parity effect, above which mode coalescence generates purely imaginary zero modes. Furthermore, by extending the lattice to include a tunable auxiliary element, we mapped the phase boundaries of higher-order singularities, identifying the exact conditions for the emergence of third-order exceptional points.

\section*{ACKNOWLEDGMENTS}

The author acknowledges the financial support of CNPq (Conselho Nacional de Desenvolvimento Cient\'ifico e Tecnol\'ogico), CAPES (Coordenação de Aperfeiçoamento de Pessoal de Nível Superior) and FAPEAL (Fundação de Amparo à Pesquisa do Estado de Alagoas).

\bibliography{apssamp}

\end{document}